# Locally maximally entangleable states and W-type states

Ri Qu, Juan Wang, Yan-ru Bao, and Zong-shang Li

*School of Computer Science and Technology, Tianjin University, Tianjin, 300072, China and*

*Tianjin Key Laboratory of Cognitive Computing and Application, Tianjin, 300072, China*

We investigate the relationship between locally maximally entangleable (LME) states and W-type states which are equivalent to a W state under stochastic local operations and classical communication (SLOCC). We prove that (i) some special W-type states of three qubits are LME; and (ii) no W-type state of four and more qubits is LME. Our results show that the W state of four and more qubits cannot be prepared by using LME states under SLOCC. However, for three qubits some special LME states can be used to prepare the W state under SLOCC.

PACS number(s): 03.67.Mn, 03.67.Ac

## I. INTRODUCTION

It is well known that there are two inequivalent classes of three-qubit genuinely entangled pure states under stochastic local operations and classical communication (SLOCC) [1]. These two classes are respectively equivalent to the *GHZ state* $\frac{1}{\sqrt{2}}(|000\rangle+|111\rangle)$ and the *W state* $\frac{1}{\sqrt{3}}(|100\rangle+|010\rangle+|001\rangle)$ under SLOCC. The W state has interesting property that its entanglement exhibits maximum robustness against the loss of one qubit. The W state can be generalized for $n$ qubits ($n \geq 3$), that is, the generalized W state is defined as

$$|W_n\rangle \equiv \frac{1}{\sqrt{n}}(|10...0\rangle+|01...0\rangle+...+|00...1\rangle). \tag{1}$$

Moreover, *W-type states* [1-4] are defined as the ones which are equivalent to the state $|W_n\rangle$ under SLOCC. Ref. [5] introduces a different class of quantum states, called locally maximally entangleable (LME) states. An approach for encoding weighted hypergraphs into (up to local unitary transformations) LME states is shown in Ref. [6].

Ref. [6] shows that the W state $|W_n\rangle$ is not LME, which implies that $|W_n\rangle$ can not be prepared by using any LME state under local unitaries. Then one may ask whether the state $|W_n\rangle$ can be prepared by some LME state under SLOCC. The main aim of this work is to answer the above question. For this, we investigate the relationship between LME states and W-type states. We show that (i) some special W-type states of three qubits are LME; and (ii) no W-type state of four and more qubits is LME. This means that our answer about the above question is "yes" for three qubits while it is "no" for four and more qubits.

This paper is organized as follows. In Sec. II, we recall notations of the trace decomposition, LME states, etc. In Sec. III, we give our main results and show the relationship between LME states and W-type states. Section IV contains our conclusions.

## II. PRELIMINARIES

Let us denote the $2\times 2$ identity matrix by $I$ and let

$$X \equiv \begin{bmatrix} 0 & 1 \\ 1 & 0 \end{bmatrix}, \quad Y \equiv \begin{bmatrix} 0 & -i \\ i & 0 \end{bmatrix}, \text{ and } Z \equiv \begin{bmatrix} 1 & 0 \\ 0 & -1 \end{bmatrix}. \tag{2}$$

Denote an operator $V$ acting on qubit $l$ by $V_l$ while $V^k$ denotes the $k$th power of the operator $V$ with $V^0 \equiv I$ for any operator $V$. Let $|\phi\rangle$ and $|\varphi\rangle$ be two pure states of $n$ qubits. We say that they are *LU equivalent* if there exist local unitary operators $\{U_l\}_{l=1,2,\ldots,n}$ such that

$$|\phi\rangle = U_1 \otimes U_2 \otimes \ldots \otimes U_n |\varphi\rangle, \tag{3}$$

i.e., $|\phi\rangle$ and $|\varphi\rangle$ are equivalent under local unitary transformations. We also say that they are *SLOCC equivalent* if there exist local invertible operators $\{A_l\}_{l=1,2,\ldots,n}$ such that

$$|\phi\rangle = A_1 \otimes A_2 \otimes \ldots \otimes A_n |\varphi\rangle, \tag{4}$$

i.e., $|\phi\rangle$ and $|\varphi\rangle$ are equivalent under SLOCC.

Let $|\phi\rangle$ be an $n$-qubit state with single qubit reduced states $\{\rho_l \equiv \text{Tr}_{\text{all but } l}(|\phi\rangle\langle\phi|)\}_{l=1,2,\ldots,n}$. For any $l$, we can write the spectral decomposition of $\rho_l$, i.e.,

$$\rho_l = U_l^\dagger D_l U_l \tag{5}$$

where $D_l = diag(\lambda_1^{(l)}, \lambda_2^{(l)})$ and $\lambda_1^{(l)} \geq \lambda_2^{(l)} \geq 0$. We call $U_1 \otimes U_2 \otimes \ldots \otimes U_n |\phi\rangle$ a *trace decomposition* of $|\phi\rangle$ [5].

Let $|\phi\rangle$ be a pure state of $n$ qubits. These $n$ qubits are called *system* ones. For each system qubit $l$ one can introduce a local *auxiliary* one $l_a$ with the initial state $|+\rangle \equiv \frac{1}{\sqrt{2}}(|0\rangle + |1\rangle)$. Let $C_l = \sum_{j=0}^{1} U_l^j \otimes |j\rangle_{l_a}\langle j|$ where $U_l$ is a unitary operator acting on system qubit $l$ and $|j\rangle_{l_a}\langle j|$ is the projector acting on the auxiliary qubit $l_a$ attached to $l$. If there exist local control gates $\{C_l\}_{l=1,2,...,n}$ such that the state $C_1 \otimes C_2 \otimes ... \otimes C_n |\phi\rangle |+\rangle^{\otimes n}$ is a maximally entangled state between the system and the auxiliary systems, then the state $|\phi\rangle$ is called *locally maximally entanglable* [5].

### III. MAIN RESULTS

In this section, we give our main results: (i) some special W-type states of three qubits are LME; and (ii) No W-type state of four and more qubits is LME. First, let us prove the result (ii) as follows.

*Proposition 1.* For $n \geq 4$, no W-type state of $n$ qubits is LME.

*Proof.* Let $|\phi\rangle$ be a W-type state of $n$ qubits. Assume that $|\phi\rangle$ is LME. It is well known that the state $|\phi\rangle$ is LU equivalent to

$$|\phi'\rangle = \sqrt{x}|00...0\rangle + \sqrt{c_1}|10...0\rangle + \sqrt{c_2}|01...0\rangle + ... + \sqrt{c_n}|00...1\rangle \qquad (6)$$

where $x \geq 0$ and $c_l > 0$ for $l \in \{1, 2, ..., n\}$ [2, 3]. Thus $|\phi'\rangle$ is also LME [7]. According to Lemma 1 in Ref. [5], there exists for each qubit $l$ a unitary operation $U_l$ such that the set $\{U_1^{l_1} \otimes U_2^{l_2} \otimes ... \otimes U_n^{l_n} |\phi'\rangle\}_{l_1,l_2,...,l_n=0,1}$ forms a normal orthogonal basis.

Let $\rho_l \equiv \text{Tr}_{\text{all but } l} (|\phi'\rangle\langle\phi'|)$. It is clear that

$$\rho_l = \begin{bmatrix} x + \sum_{r \neq l} c_r & \sqrt{xc_l} \\ \sqrt{xc_l} & c_l \end{bmatrix}, \qquad (7)$$

which implies $|\phi'\rangle$ is one trace decomposition if and only if $x = 0$. From (7), it is clear that

$$\det(\rho_l) = c_l \sum_{r \neq l} c_r \qquad (8)$$

According to (5) and (8), it is known that $D_l \propto I$ if and only if $x = 0$ and $c_l = \frac{1}{2}$. This implies that for $n \geq 4$ there is at most one qubit $t \in \{1, 2, ..., n\}$ such that $D_t \propto I$. Then there are only two cases which should be discussed as follows. (a) $\forall l \in \{1, 2, ..., n\}, D_l \not\propto I$. It is known that there is a real number $\alpha_l$ such that

$$U_l = R_{Z_l}(\alpha_l) X_l R_{Z_l}(-\alpha_l) = \begin{bmatrix} 0 & e^{i\alpha_l} \\ e^{-i\alpha_l} & 0 \end{bmatrix} \qquad (9)$$

where $R_{Z_l}(\alpha_l) \equiv e^{i\alpha_l Z_l/2}$ [5]. For two different qubits $j$ and $k$, we can obtain

$$\langle \phi'|U_j \otimes U_k|\phi'\rangle = 2\sqrt{c_j c_k}\cos(\alpha_j - \alpha_k) \qquad (10)$$

It is impossible that $\cos(\alpha_j - \alpha_k) = 0$ for any two qubits. In fact, assume that $\cos(\alpha_j - \alpha_k) = 0$ and $\cos(\alpha_k - \alpha_l) = 0$. Then we would obtain $|\cos(\alpha_j - \alpha_l)| = 1$.

(b) There exists a unique quit $t \in \{1, 2, ..., n\}$ such that $D_t \propto I$.

Let $j, k, l \in \{1, 2, ..., n\} - \{t\}$ be three different qubits. Then we can obtain that $D_j \not\propto I$, $D_k \not\propto I$ and $D_l \not\propto I$. Similar for (a), it is known that at least one of three states, i.e., $U_j \otimes U_k |\phi'\rangle$, $U_k \otimes U_l |\phi'\rangle$ and $U_j \otimes U_l |\phi'\rangle$, is not orthogonal to $|\phi'\rangle$. ∎

According to the above proposition, it is known that the W state of four and more qubits cannot be prepared by using any LME state under SLOCC. However, the following proposition implies that for three qubits some special LME states can be converted into the W state by means of SLOCC.

*Proposition 2.* Some special three-qubit W-type states are LME.

*Proof.* Suppose that a W-type state $|\phi\rangle$ of three qubits is LU equivalent to

$$|\phi'\rangle = \sqrt{c_1}|100\rangle + \sqrt{c_2}|010\rangle + \sqrt{\frac{1}{2}}|001\rangle \qquad (11)$$

where $c_1 > 0$, $c_2 > 0$, and $c_1 + c_2 = \frac{1}{2}$. We define the local unitary operators as

$$U_l = \begin{cases} R_{Z_l}(\alpha_l) X_l R_{Z_l}(-\alpha_l) & l = 1, 2 \\ Z_l & l = 3 \end{cases} \qquad (12)$$

where the real numbers $\alpha_1, \alpha_2$ which satisfy $\cos(\alpha_1 - \alpha_2) = 0$. Clearly, for $l = 1, 2, 3$ the operator $U_l$ is both unitary and hermitian. Then we can obtain that

$$U_1|\phi'\rangle = \sqrt{c_1}e^{i\alpha_1}|000\rangle + \sqrt{c_2}e^{-i\alpha_1}|110\rangle + \sqrt{\frac{1}{2}}e^{-i\alpha_1}|101\rangle,$$

$$U_2|\phi'\rangle = \sqrt{c_1}e^{-i\alpha_2}|110\rangle + \sqrt{c_2}e^{i\alpha_2}|000\rangle + \sqrt{\frac{1}{2}}e^{-i\alpha_2}|011\rangle,$$

$$U_3|\phi'\rangle = \sqrt{c_1}|100\rangle + \sqrt{c_2}|010\rangle - \sqrt{\frac{1}{2}}|001\rangle,$$

$$U_1 \otimes U_2 |\phi'\rangle = \sqrt{c_1} e^{i(\alpha_1-\alpha_2)} |010\rangle + \sqrt{c_2} e^{-i(\alpha_1-\alpha_2)} |100\rangle + \sqrt{\frac{1}{2}} e^{-i(\alpha_1+\alpha_2)} |111\rangle,$$

$$U_1 \otimes U_3 |\phi'\rangle = \sqrt{c_1} e^{i\alpha_1} |000\rangle + \sqrt{c_2} e^{-i\alpha_1} |110\rangle - \sqrt{\frac{1}{2}} e^{-i\alpha_1} |101\rangle,$$

$$U_2 \otimes U_3 |\phi'\rangle = \sqrt{c_1} e^{-i\alpha_2} |110\rangle + \sqrt{c_2} e^{i\alpha_2} |000\rangle - \sqrt{\frac{1}{2}} e^{-i\alpha_2} |011\rangle,$$

$$U_1 \otimes U_2 \otimes U_3 |\phi'\rangle = \sqrt{c_1} e^{i(\alpha_1-\alpha_2)} |010\rangle + \sqrt{c_2} e^{-i(\alpha_1-\alpha_2)} |100\rangle - \sqrt{\frac{1}{2}} e^{-i(\alpha_1+\alpha_2)} |111\rangle. \quad (13)$$

According to (13), it is clear that the state set $\{U_1^{l_1} \otimes U_2^{l_2} \otimes U_3^{l_n} |\phi'\rangle\}_{l_1,l_2,l_3=0,1}$ forms a normal orthogonal basis. ∎

## VI. CONCLUSIONS

The W state $|W_n\rangle$ is one of famous $n$-partite genuinely entangled pure states of $n$ qubits. Since it has been applied for several quantum information processing tasks [8], the preparation of the W state is very important. Ref. [6] show the state $|W_n\rangle$ cannot be prepared by using LME states (including graph states [9] and hypergraph states [10]) under local unitaries. In this paper, we investigate the relationship between LME states and W-type states in order to answer whether the W state can be prepared by means of some LME state under SLOCC. The relationship is shown as follows: (i) some special W-type states of three qubits are LME; and (ii) no W-type state of four and more qubits is LME. Therefore, it is interesting that there are two different answers for the above question according to the number of qubits. For three qubits, the W state can be prepared by using some special LME states under SLOCC. However, the W state of four and more qubits cannot be converted into any LME state under SLOCC.

## ACKNOWLEDGMENTS

This work is supported by the Chinese National Program on Key Basic Research Project (973 Program, Grants No. 2013CB329301 and No. 2013CB329304) and the Natural Science Foundation of China (Grants No. 61170178, No. 61105072 and No. 61272265). This work is completed during our academic visit at the Department of Computing, the Open University, U. K..